\documentclass [prb, superscriptaddress, showpacs, preprint]{revtex4}

\usepackage{graphicx}

\newcommand{\grad}{^\circ}
\newcommand{\srmno}{SrMnO$_3$}
\newcommand{\rdmno}{R$_{1-x}$D$_x$MnO$_3$}

\newcommand{\mnt}{Mn$^{3+}$}
\newcommand{\mnq}{Mn$^{4+}$}
\newcommand{\tn}{$T_N$}

\newcommand{\mnoct}{MnO$_6$}

\begin{document}

\title{Structural phase transition and magnetism in hexagonal \srmno}

\author{A. Daoud-Aladine}
\affiliation{ISIS facility, Rutherford Appleton Laboratory-CCLRC,
Chilton, Didcot, Oxfordshire, OX11 0QX, United Kingdom. }
\author{C. Martin}
\affiliation{Laboratoire CRISMAT-UMR, 6508 ENSI CAEN, 6, Marechal
Juin, 14050 Caen, France} \affiliation{ISIS facility, Rutherford
Appleton Laboratory-CCLRC, Chilton, Didcot, Oxfordshire, OX11 0QX,
United Kingdom. }
\author{L. C. Chapon}
\affiliation{ISIS facility, Rutherford Appleton Laboratory-CCLRC,
Chilton, Didcot, Oxfordshire, OX11 0QX, United Kingdom. }
\author{M. Hervieu}
\affiliation{Laboratoire CRISMAT-UMR, 6508 ENSI CAEN, 6, Marechal
Juin, 14050 Caen, France}
\author{K. S. Knight}
\affiliation{ISIS facility, Rutherford Appleton Laboratory-CCLRC,
Chilton, Didcot, Oxfordshire, OX11 0QX, United Kingdom. }
\author{M. Brunelli}
\affiliation{European Synchrotron Radiation Facility, BP220, F-38043
Grenoble Cedex, France}
\author{P.G. Radaelli}
\affiliation{ISIS facility, Rutherford Appleton Laboratory-CCLRC,
Chilton, Didcot, Oxfordshire, OX11 0QX, United Kingdom. }
\affiliation{Dept. of Physics and Astronomy, University College
London, Gower Street, London WC1E 6BT, United Kingdom}
\date{\today}

\begin{abstract}
The structural and magnetic properties of the hexagonal four-layer
form of \srmno\ have been investigated by combining magnetization
measurements, electron diffraction and high-resolution synchrotron
X-ray and neutron powder diffraction. Below 350K, there is subtle
structural phase transition from hexagonal symmetry (space group
$P6_3/mmc$) to orthorhombic symmetry (space group $C222_1$) where
the hexagonal metric is preserved. The second-order phase transition
involves a slight tilting of the corner-sharing Mn$_{2}$O$_{9}$
units composed of 2 face-sharing MnO$_6$ octahedra and the
associated displacement of Sr$^{2+}$ cations. The phase transition
is described in terms of symmetry-adapted displacement modes of the
high symmetry phase. Upon further cooling, long range magnetic order
with propagation vector $\mathbf{k}=(0,0,0)$ sets in  below 300K.
The antiferromagnetic structure, analyzed using representation
theory, shows a considerably reduced magnetic moment indicating the
crucial role played by direct exchange between Mn centers of the
Mn$_{2}$O$_{9}$ units.

\end{abstract}


\maketitle

\section{Introduction}

Transition metal oxides show remarkable physical properties, such as
colossal magnetoresistance in manganites\cite{Dagotto01}, or
superconductivity in cuprates\cite{Orenstein00}, which after decades
of researches, remain only partly understood. This reflects our
understanding of the structural, electronic and magnetic phenomena,
which is well established only in the limit where the systems show
localized or itinerant electron behavior\cite{Imada98}. Doping with
electrons or holes is the canonical method to explore the
intermediate regime.  For example, in \rdmno \ manganese perovskites
(R:trivalent cation, D: divalent cation), the Mn valence is formally
intermediate 3+x, affecting the electrical conductivity and leading
to a complex phase diagram.

However, even undoped compounds display an intriguing evolution of
their magnetic properties as the geometry and coordination are
changed.  For example, in the RMnO$_3$ and DMnO$_3$ end members of
the \rdmno \ series,  Mn has undoubtedly well localized electrons,
making these compounds prototype antiferromagnetic (AF) Mott
insulators. Their magnetic properties fit a Heisenberg picture, in
which superexchange (SE) interactions couple high spin \mnt$(S=2)$
and \mnq $(S=3/2)$ ions at $x=0$ and $x=1$
respectively\cite{goodenough55,Anderson59,Kanamori59}. This can be
contrasted to the very unusual magnetic properties of some hexagonal
manganese halides containing single valent Mn ions.
CsMnBr$_3$\cite{Mason89} and CsMnI$_3$\cite{Harrison91} compounds
containing S=5/2 high spin Mn$^{2+}$ ions arranged in infinite
strings of face-sharing \mnoct \ octahedra parallel to the c axis,
from which exchange frustration is expected between the direct and
the super-exchange interactions of different sign within the chains.
Exchange frustration in the basal plane also bring about chirality
of the magnetic state and a spin dynamic corresponding to new
universality classes\cite{Mason89}. One may be tempted to conclude
that these unusual magnetic properties arise from the quasi
one-dimensional nature of the exchange, combined with easy axis
anisotropy.  However, isostructural BaMnO$_3$ (S=3/2) does not show
such effects, but it has instead a simple antiferromagnetic
structure and rather classical magnetic behavior with the ordered
moment value saturating at low temperatures\cite{Christensen72}.

In order to separate the influence upon the magnetic properties of
the lattice geometry from more conventional chemical effects, the
study of polymorphic compounds, which display different stable or
metastable structures with the same chemical composition is
particularly valuable.  As discussed by Nagas and
Roth\cite{Negas70}, the stability of the structures adopted by
$ABX_3$ compounds depends on the tolerance factor
$t=(r_A+r_X)/\sqrt{2}(r_{B}+r_X)$, where $r_A$ is the average ionic
radius of the A-site cations, while $r_{B}$and $r_X$ are those of
the B-site metal and of the anion (oxygen, halogen or chalcogen).
Most manganites have $t\leq 1$, and crystallize in orthorhombic or
rhombohedral perovskite structures showing cooperative tilting of
the corner sharing \mnoct \ array, with respect to the ideal cubic
perovskite. For sufficiently large $r_A$, $t>1$ materials adopt a
different hexagonal symmetry, as in the case of BaMnO$_3$ and the
aforementioned halides. For $t\gtrsim 1$, polytypes are obtained
depending on the synthesis procedure\cite{Syono69,Negas70}. These
polytypes are characterized by the stacking sequence of the AO$_3$
layers, which controls how the \mnoct \ octahedra arrange. Octahedra
are sharing corners across the central AO$_3$ layer of
\emph{..abc..} sequences, while they share faces across
\emph{..aba..} type sequences.

\begin{figure}[tbp]
  \includegraphics[width=400pt]{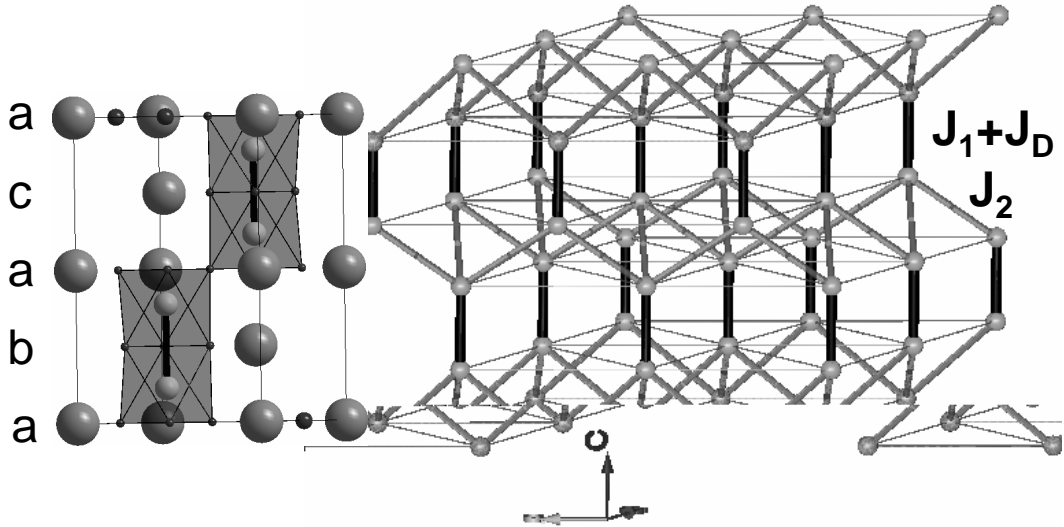}\\
  \caption{Left panel: Crystal structure of \srmno \ viewed along the [110] direction. The Sr, Mn and O atoms are shown as large grey spheres,
  small grey spheres and black spheres respectively. Face-sharing MnO$_6$ octahedra are also represented as transparent grey units;
  Right panel: magnetic exchange topology of the Mn sublattice. Black bonds represent magnetic exchange between Mn ions of the Mn$_2$O$_9$
  units, composed of antiferromagnetic direct exchange (J$_D$) and ferromagnetic super-exchange (J$_1$). Grey bonds represent super-exchange interactions (J$_2$) between Mn ions
  of the hexagonal units. See text for details.}
  \label{magtopo}
\end{figure}

For example, synthesized at ambient pressure in air, \srmno \
($t=1.05$) has a four layer (4L) \emph{abac} stacking sequence of
SrO$_3$ layers as shown in Fig.\ref{magtopo}, whereas  modified
synthesis routines are required to obtain the cubic
perovskite\cite{Chmaissem01}. 4L is a rather rare polytype among
ABX$_3$ compounds, and represents an intermediate case between the
hexagonal (2L) polytype adopted by BaMnO$_3$, and the familiar
perovskite structure of CaMnO$_3$ containing only corner sharing
octahedra. It is therefore interesting to study the magnetism in
\srmno \, in comparison with that of the perovkite and 2L hexagonal
materials. This could also help to shed light on the striking
differences of magnetic behavior in isostructural CsMnBr$_3$ and
BaMnO$_3$. In \srmno, a complex magnetic behavior is expected from
the exchange topology, which results from the arrangement of pairs
of face sharing \mnoct \ linked connected by the octahedra corners.
In this structure, direct Mn-Mn exchange interactions between \mnq \
ions in face sharing octahedra ($J_D$) compete with the $90\grad$
AF-SE linkages through the face shared by \mnoct \ octahedra ($J_1$)
as in BaMnO$_3$, and $180\grad$ AF-SE linkages across the common
oxygen of corner sharing \mnoct \ octahedra ($J_2$), like in
CaMnO$_3$. As in several hexagonal transition metal oxides or
sulfides, $J_D$ is probably very strong, since the metal-metal
distance is much lower than that observed in the corresponding
intermetallic compounds containing the same transition metal
element\cite{Marasinghe02,Bai84}.

\begin{figure}[tbp]
  \includegraphics[width=400pt]{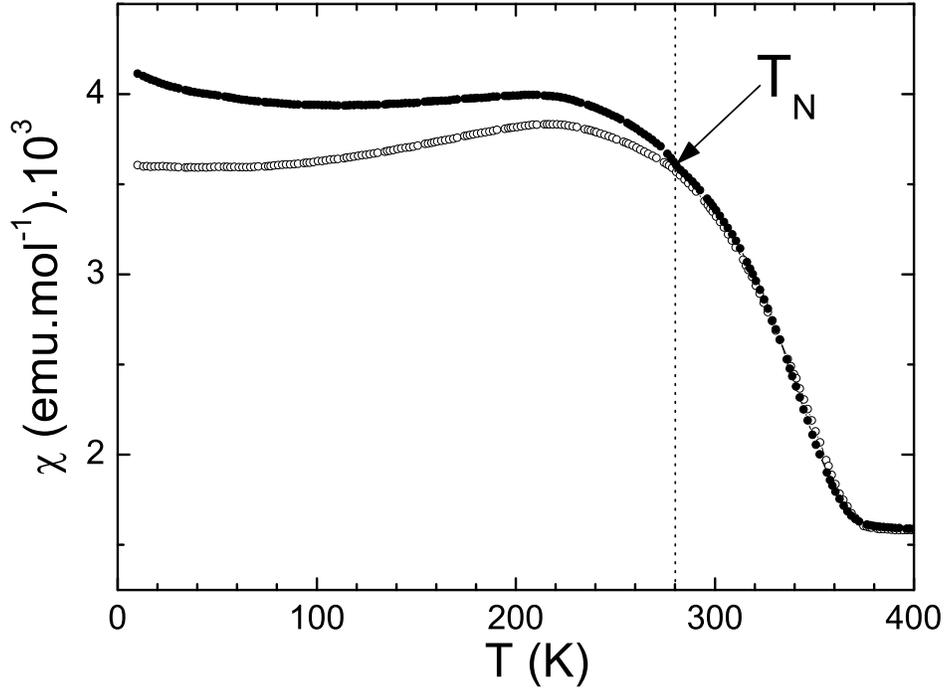}\\
  \caption{DC Magnetic susceptibility of \srmno \ under a magnetic field of 1000 Oe. Data in zero-field cooled (grey symbols) and field cooled (black) conditions are shown.}
  \label{ppms}
\end{figure}

As first noticed by Battle \emph{et al.}\cite{Battle88}, \srmno \
shows unusual magnetic properties: the magnetization data, which we
have also measured (see fig.\ref{ppms}) show a transition above room
temperature ($T_s\sim 350K$), but neutron powder diffraction (NPD)
demonstrates that this does not correspond to the AF ordering. AF
order sets in only at lower temperatures, where zero field cooled
and field cooled $M(T)$ curves deviate (\tn$=278K$)\cite{Battle88}.
Recently, Raman spectroscopy was used to evidence a structural
transformation in \srmno \ below $200K$\cite{sacchetti05}. To
explain the subtle difference observed by Raman spectroscopy,  the
authors proposed a 4L to 6L transition, allowing the hexagonal
symmetry to be retained.

In this article, we present a complete study of the magnetic and
structural properties of \srmno, using in combination electron
microscopy, neutron powder diffraction (NPD) and high-resolution
synchtrotron X-ray powder diffraction (HR-SXPD).

\section{Experimental}

Polycristalline samples of \srmno\ were synthesized by solid-state
reaction in air of a stoichiometric mixture (1:0.5) of SrCO$_3$ and
Mn$_2$O$_3$. The powder was heated at 950$\grad$C during two weeks
with intermediate grindings, then pelletized and sintered at
1000$\grad$C for 12 hrs. This low-temperature synthesis was
determined from the phase diagrams \cite{Syono69,Negas70} as the
best route to obtain pure hexagonal 4L-structure without oxygen
deficiency. Samples quality and stoichiometry were determined by
preliminary neutron diffraction measurements.

At room temperature, the reconstruction of the reciprocal space was
carried out with a JEOL 200 CX electron microscope and high
resolution electron microscopy (HREM) images at room temperature
were recorded using a TOPCON 002B microscope (Cs = 0.4mmm). The
electron diffraction patterns versus temperature were collected with
a JEOL 2010 electron microscope operating at 200kV and equipped with
a double-tilt liquid N2 sample holder (tilt $\pm30\grad$,
$\pm45\grad$, and $90K<T300K$). All the patterns were recorded under
the same experimental conditions, i.e. the same exposure time, the
same electron beam alignment and the same beam intensity and
increasing the temperature from $90K$ to $300K$, in steps of $10$
degrees, waiting for temperature stabilization before recording. The
three microscopes are equipped with Energy Dispersive Analysis
spectrometers (EDS), which allow determining the Sr/Mn ratio.

Magnetic susceptibility was collected under a magnetic field of 1000
Oe under zero-field cooled (ZFC) and field cooled (FC) processes.
Data were collected on warming between 2K and 400K at a sweep rate
of 1K/min, using a Vibrating Sample Magnetometer (Quantum Design,
PPMS).

High resolution synchrotron and time-of-flight neutron powder
diffraction data were collected at 100 and 350K, using the ID31 beam
line at the ESRF, Grenoble with a wavelength $\lambda =
0.30001(2)\AA$, and HRPD at the ISIS Facility (UK), respectively.
Subsequent medium-resolution neutron diffraction data were collected
on the GEneral Materials (GEM) Diffractometer of the ISIS facility
to study the temperature dependence of the crystal structure and the
magnetic structure. Data sets at temperature between $1.9$ and
$300K$ with $25K$ steps were collected on warming using a standard
He cryostat. The diffraction patterns were used to refine the \srmno
\ structure by the Rietveld method\cite{Rietveld69} using the
FullProf suite (Ref.\onlinecite{Rodriguez-Carvajal93}). Symmetry
analysis for the structural and magnetic phase transitions was
performed using the BasIreps program, part of the FullProf suite as
well as programs available on the Bilbao Crystallographic Server
\cite{Aroyo06}.

\section{Results}

\subsection{Magnetization measurements}
The DC magnetic susceptibility of \srmno\ between 2 and 400K is
reported in fig.2. Plot of the inverse suceptibility, not shown,
indicates that a Curie-Weiss behavior is not obeyed in this
temperature range, in agreement with earlier results
\cite{Chamberland70}. However, our results below 400K are
qualitatively different to what was previously reported. Below 380K,
there is a clear upturn in the susceptibility followed by a
relatively shallow decrease below 300K. The upturn could be
associated with the structural phase transition reported herein, and
discussed in the following sections, as evidenced by X-ray and
neutron diffraction. However, it was impossible to track precisely
the structural details as a function of temperature due to the
extremely weak intensities of the superlattice reflections and
therefore confirm unambiguously the correlation between structural
and magnetic behaviors. Also, this upturn could be associated to
short-range magnetic correlations within the
face-sharing MnO$_6$ octahedra, as suggested by Battle et al.\cite{Battle88}.\\
At lower temperatures (T$<$300K) where $\chi$ decreases, the
behavior is clearly reminiscent of antiferromagnets. This is in
agreement with previous work, in particular with the largely
negative Weiss temperature of -1210K derived by susceptibility
measurements at high temperature\cite{Chamberland70}. The
determination of the N\'{e}el temperature from susceptibility alone
is not straightforward due to the broad maximum observed in the
variation of $\chi$. Battle and co-workers identified a N\'{e}el
temperature of 278K from Mossbauer spectroscopy on Fe-doped
samples\cite{Battle88}. Our measurements suggest the same N\'{e}el
transition temperature since the susceptibility curves in field
cooled (FC) and zero-field cooled (ZFC) conditions slightly deviates
at around 280K. Moreover, this accident coincides with the onset of
a long-range ordered antiferromagnetic structure evidenced by our
neutron diffraction experiment, as discussed below.

\subsection{Crystal Structure}
In this section, we report a detailed crystallographic
characterization of \srmno\ from electron microscopy and
high-resolution X-ray and neutron powder diffraction experiments.

\subsubsection{Transmission Electron Microscopy}

\begin{figure}[tbp]
 \includegraphics[width=400pt]{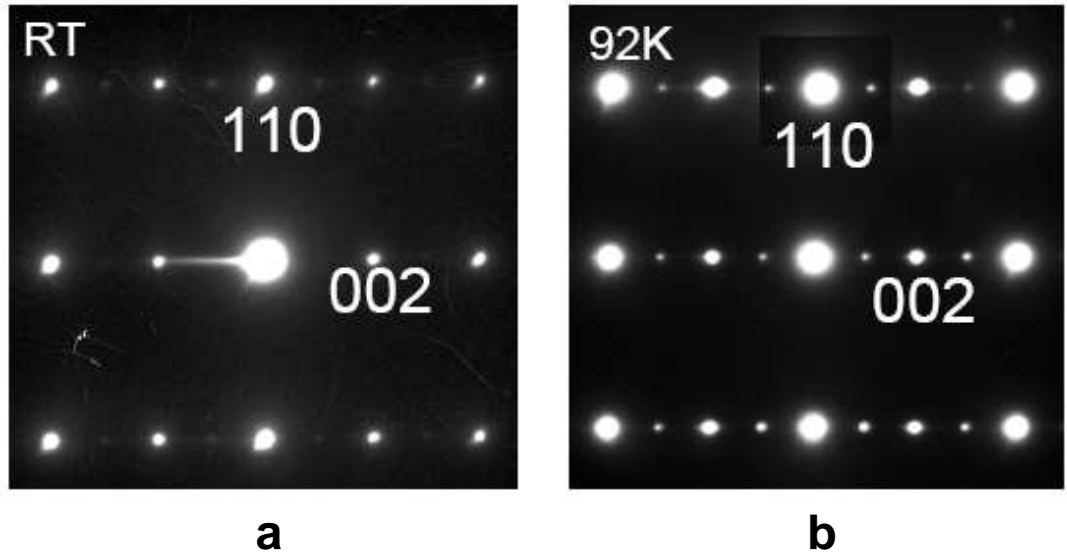}\\
  \caption{$[1\bar{1}0]$ ED pattern of \srmno \ at RT (a) and $92K$ (b). The low temperature data shows
  the appearance of another system of weak reflections,  \emph{hh2$\bar{h}$l l = no
condition}, which clearly violate the mirror $c$. A careful rotation
around $\vec{c}^*$, selecting the $[hk0]$ ED patterns with large $h$
and $k$ values, shows that, in our conditions of reflection, the
conditions \emph{00l: l=2n+1} are scarcely visible but not null and
hence, are not due to double diffraction.}
  \label{tem1}
\end{figure}

A large number of crystallites have been analyzed by EDS and
electron diffraction, confirming an homogeneous Sr/Mn ratio =1 in
the limit of the accuracy of the technique. The reconstruction of
the reciprocal space was carried out by tilting the sample around
the crystallographic axes. It evidences an hexagonal cell with
$a\approx5.45\AA\approx \sqrt{2}a_p$ ($a_p$ is the parameter of the
ideal cubic perovskite cell), $c\approx9.06\AA$ and the extinction
condition \emph{hh2$\bar{h}$l =2n}. These results are consistent
with the 4L polytype proposed for \srmno \ as well as the space
group $P6_3/mmc$ previously reported\cite{Syono69,Negas70}. The
$[1\bar{1}0]$ ED pattern at room temperature is given in
Fig.\ref{tem1}a. The quality of the crystallites was checked using
HREM, in order to detect the presence of any intergrowth defects
(variations in the layer stacking along $\vec{c}$ , i.e. the
presence of different polytypes). The lattice images, not shown,
confirm that the stacking of the 4L structure is quite perfect. The
density of defects is very low (of the order a few events for a
nanometer-sized crystallite), which is rather rare in the hexagonal
polytypes, allowing one to conclude that these defect do not significantly affect the symmetry.\\
The reciprocal space was reconstructed at $90K$, tilting especially
around the $\vec{c}^*$  axis. One clearly observes two important
points : first, the lattice parameters remains unchanged, especially
$c^*$, attesting that the 4L stacking mode is retained and second,
the appearance of another system of weak reflections,
\emph{hh2$\bar{h}$l l = no condition}, which clearly violate the
mirror $c$ symmetry. This is illustrated comparing Fig.\ref{tem1}a
and \ref{tem1}b with the evolution of the $[1\bar{1}0]$ ED pattern
between RT and $92K$. At this step, a careful rotation around
$\vec{c}^*$, selecting the $[hk0]$ ED patterns with large $h$ and
$k$ values, shows that, in our conditions of reflection, the Bragg
peaks  \emph{00l: l=2n+1} are scarcely visible but not null.
Increasing $T$, step by step, up to room temperature shows that the
intensity of the extra reflections remain unchanged between $90K$
and $200K$, it decreases from $200K$ to $265K$ and, lastly,
disappears for $T>267K$ when keeping our working conditions
constant. This transition temperature is probably an underestimate,
due to the heating of the sample under electron beam irradiation.

\subsubsection{Combined NPD and HR-SXPD refinements at $T=350K$ and $T=100K$}

\begin{figure}[tbp]
  \includegraphics[width=400pt]{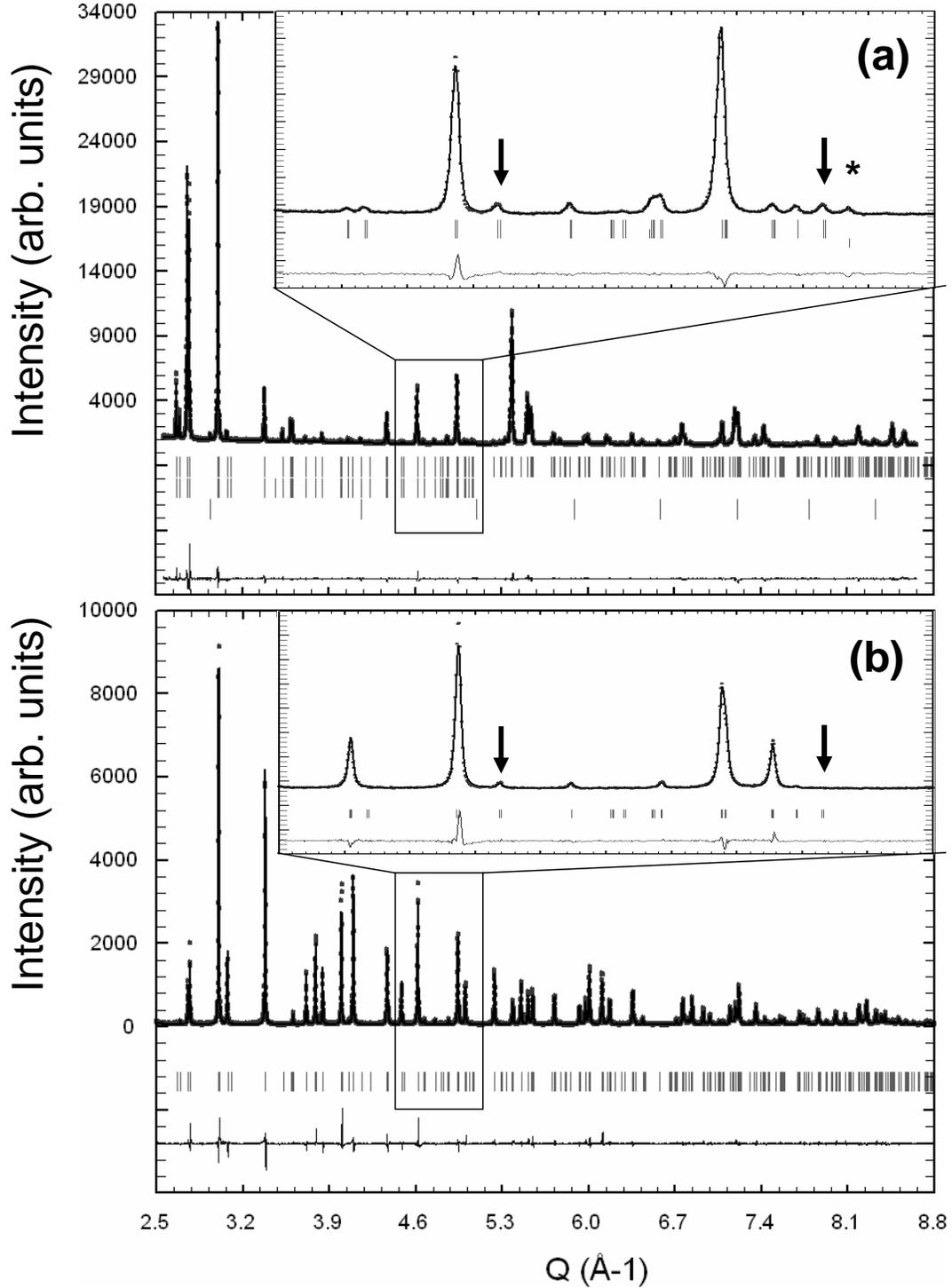}\\
  \caption{Combined Rietveld refinement of the \srmno \ structure at 100K with (a) NPD HRPD and (b) HR-SXPD ID-31 data collected on \srmno \ at $T=100K$.
  For both patterns, the continuous line correspond to the model fitting to the data, the lowest continuous line, the difference curve,
  and ticks are indicating the position of the Bragg reflections. For the NPD (a), the upper, middle and lower ticks
  represents the contribution from the nuclear scattering, the magnetic scattering, and the contribution from the Vanadium can (asterisk).
  Arrows indicate the position of the few reflections violating the extinction conditions of the high temperature space group $P6_3/mmc$.}
  \label{combi}
\end{figure}
\begin{figure}[tbp]
  \includegraphics[width=440pt]{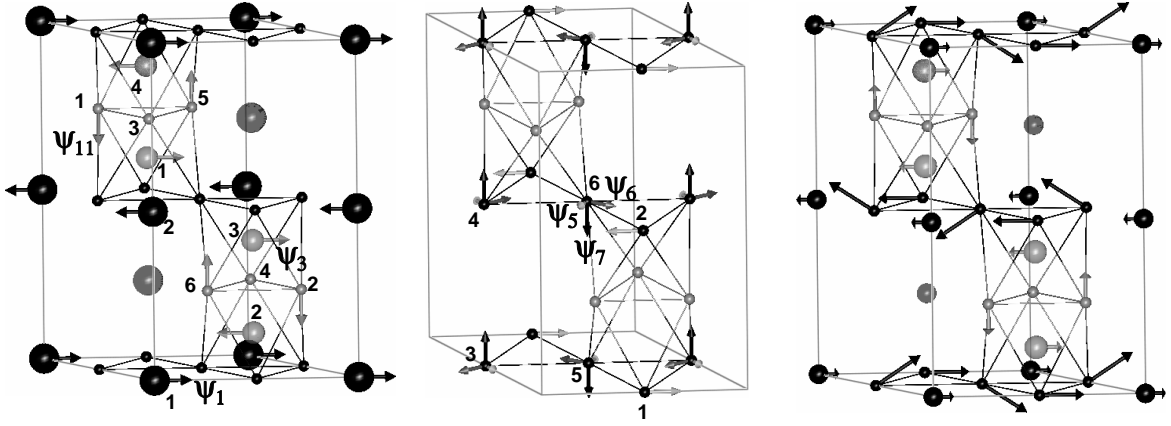}\\
  \caption{Atomic displacement modes responsible for the structural phase transition in \srmno. The arrows and the atom labels
  gives a graphical representation of the basis vectors of Irrep. $\Gamma_5$ involved in the transition, which are
  given table IV, Appendix A (a) Displacement modes for
  Mn (medium sized grey spheres), Sr$_1$ (large black spheres) and O$_2$ (small grey spheres) (b) Displacement modes for O$_1$
  (c) Resultant distortion from the linear combination $\sum_{\lambda} C_\lambda.\psi_\lambda$ of all basis functions: the refinement leads to $C_1>0$, $C_3<0$ and $C_{11}<0$ for Sr$_1$, Mn, and O$_2$ respectively
  For O$_1$, the mixing coefficients can be constrained to have no component along b, as $C_\alpha(\psi_{5}-\psi_{6})+C_\beta\psi_7$}
  \label{modes}
\end{figure}

In agreement with the TEM results, high resolution X-ray and neutron
powder diffraction indicate that a number of (hkl) Bragg reflections
with odd l index appear at low temperature. In the synchrotron X-ray
diffraction pattern at 100K, the strongest of these additional peaks
is (221) whereas in the neutron data, two reflections indexed (221)
and (223) in the hexagonal setting are the most clearly visible. The
fact that these additional reflections appear at high value of the
scattering vector (Q) and are present in the X-ray data, indicate
their nuclear origin. At 350K, the reflections are almost extinct
but a close inspection of the diffraction pattern reveal a weak
contribution even at this temperature. Nevertheless, their
intensities at 350K are too weak to determine the change of
structure. In this respect, the model of Battle et al., with
hexagonal symmetry P6$_3$/mmc, remains a valid approximation at high
temperature. The extracted parameters from the combined refinement
of X-ray and
neutron data at 350K, assuming P6$_3$/mmc symmetry, are reported in table \ref{TableCombi}. \\
At low temperature, this model is obviously invalid due to the
presence of new reflections. Also, our preliminary study with medium
resolution neutron diffraction as a function of temperature,
indicates that the thermal parameter of some atoms, in particular,
O(1), increases when the temperature is lowered, suggesting that
P6$_3$/mmc is a pseudo-symmetry. In agreement with our TEM results,
NPD and HR-SXPD data are consistent with the absence of the
\textit{c} glide-plane symmetry operation. None of the maximal
isotropy subgroups of order 2, as listed in the International Tables
of crystallography\cite{IT83}, is able to fit the data. The solution
was found among isotropy subgroups of higher index, which are
compatible with the Landau and Lifshitz conditions of the theory of
second order phase transitions, as listed by Hatch and
Stokes\cite{Stokes84} or equivalently, by the program
Isotropy\cite{Stokes02}. The only solution fitting the data
corresponds to the orthorhombic space group $C222_1$ with a
super-cell related to the hexagonal cell in the following way:
$\mathbf{a}_o=\mathbf{a}_h$,
$\mathbf{b}_o=\mathbf{a}_h+2\mathbf{b}_h$,
$\mathbf{c}_o=\mathbf{c}_h$. This phase transition is a perfect
example of pseudo-symmetry problems since there are no direct
indication of the orthorhombic symmetry from splitting of equivalent
reflections or anisotropic broadening. The system remains metrically
hexagonal within the very high resolution of our X-ray and neutron
experiments; therefore in-plane lattice parameters were constrained
accordingly in the refinements. This also indicate that the atomic
displacements involved must be
small enough to remain decoupled from the lattice strain.\\
In the low-temperature structure, the atomic positions for all
atoms, with the exception of the Mn site, split into two orbits. In
our model, all refinable atomic positions were allowed to vary,
while thermal parameters of atoms belonging to the two related
orbits were constrained to be equal. The combined X-ray and neutron
Rietveld refinement, presented in Fig.\ref{combi}, is in excellent
agreement with the data. Reliability factors of the refinement and
refined structural parameters are listed in Table \ref{TableCombi}.
It is important to note that the combined refinement is of crucial
importance here because the difference in contrast in the X-ray and
Neutron experiments (O is a strong neutron scatterer, Mn neutron
scattering
length is negative) is extremely selective.\\
The nature of the phase transition, involving all ions in the
crystal, is difficult to comprehend without decomposing the
respective atomic displacements. We have used group theory, to
explain in details the symmetry descent across the phase transition.
The atomic displacements, shown in Fig. \ref{modes}, are decomposed
in symmetry-adapted modes of the irreducible representation
$\Gamma_5$ corresponding to the observed symmetry descent. More
details about the treatment with representation theory are given in
Appendix \ref{sa}, and the basis vectors of these modes are reported
in tables \ref{bsf}. Figs.\ref{modes}a and \ref{modes}b represent
individual modes while Fig.\ref{modes}c shows the resulting
displacements consistent with the final refinement of atomic
positions. The transition involves a slight tilting of the
corner-sharing Mn$_{2}$O$_{9}$ units composed of 2 face-sharing
MnO$_6$ octahedra. This is coupled to a shear motion of Sr$^{2+}$
cations in the \textit{ab}-plane. It is clear that the largest
displacements are associated with the O(1) ions which explain our
preliminary results of large thermal displacements on this site in
the hexagonal setting. The sequential refinements of GEM data
($1.5K<T<300K$) do not show any obvious discontinuity for most of
the geometrical parameters of the structure: excepted for volume
contraction, the Mn-O bond distances and angles within the
Mn$_{2}$O$_{9}$ units are temperature independent. The structural
transition essentially manifests in the temperature dependance of
some of the Mn-O-Mn angles across corner sharing octahedra, which
very slightly reduce by $2\grad$ in the low temperature phase. The
Raman data\cite{sacchetti05} must therefore be reinterpreted in the
light of the rather subtle transition we have established.

\begin{table}[p]
\begin{center}
\begin{tabular}[b]{*{6}{c}}
\hline \hline
\multicolumn{6}{c}{$T=350K$: $P6_3/mmmc$, $a_h=b_h=5.461(1)\AA$, $c=9.093(2)\AA$ \footnote{NPD $\chi^{2}$=8.47  $R_{Bragg}=5.23\%$ ; HR-SXPD  $\chi^{2}=6.91$  $R_{Bragg}=5.71  \%$} } \\
\hline
           &  Sr$_1 (2a)$ &   Sr$_2 (4f)$   &    Mn$(4f)$       &   O$_1(6g)$ &     O$_2 (6h)$         \\
Position   &   (0,0,0)    &   (1/3,2/3,1/4) &    (1/3,2/3,z)    &   (1/2,0,0) &   (-x,x,3/4)      \\
           &              &                 &  z=0.61264(  20)  &             & x=0.81858(  25)   \\
B$_{iso}$  &  0.67( 8)    &    0.40(8)      &    0.26(7)        &     0.69(8) &    0.40(7) \\
\end{tabular}
\\
\begin{tabular}[b]{*{6}{c}}
\hline \hline
\multicolumn{6}{c}{$T=100K$: $C222_1$, $a=5.4435(1)\AA$, $b\approx\sqrt{3}b_h=9.4211(2)\AA$, $c=9.0630(1)\AA$ \footnote{NPD $\chi^{2}$=4.37  $R_{Bragg}=2.64\%$ ; HR-SXPD  $\chi^{2}=2.51$  $R_{Bragg}=3.78 \%$} } \\
\hline
Atom   &   Wyck   &    x        &     y          &     z           &    B$_{iso}$       \\
\hline
 Sr1   &  $4a$     &  0.0096(8)  &     0          &     0           &    0.44(3)   \\
 Sr2   &  $4b$     &  0          &     1/3        &     1/4         &    0.44(3)   \\
 Mn    &  $8c$     & -0.0123(8)  &     1/3        &     0.6131(2)   &    0.33(4)   \\
 O11   &  $4a$     &  0.5212(6)  &     0          &     0           &    0.64(4)   \\
 O12   &  $8c$     &  0.7712(6)  &     1/4        &     0.0085(5)   &    0.64(4)   \\
 O21   &  $4b$     &  0          &     -0.1798(2) &     1/4         &    0.52(4)   \\
 O22   &  $8c$     &  0.2696(3)  &     0.0899(1)  &     0.2412(6)   &    0.52(4)   \\
\hline
\end{tabular}
\end{center}
\caption{Structural parameters of \srmno, obtained by combined
refinements of HRPD-ISIS, and ID31-ESRF neutron and synchrotron
data. At $100K$, structural parameters have been constrained as:
$y(Sr_2)=1/3$, $y(Mn)=1/3$, $y(O_{12})=1/4$, $x(O22)=-3/2y(O21)$ and
$x(O22)=-1/2y(O21)$. We also constrained the isotropic temperature
factors to be the same for each chemical specie} \label{TableCombi}
\end{table}

\subsection{Magnetic structure}

\begin{figure}[tbp]
  \includegraphics[width=400pt]{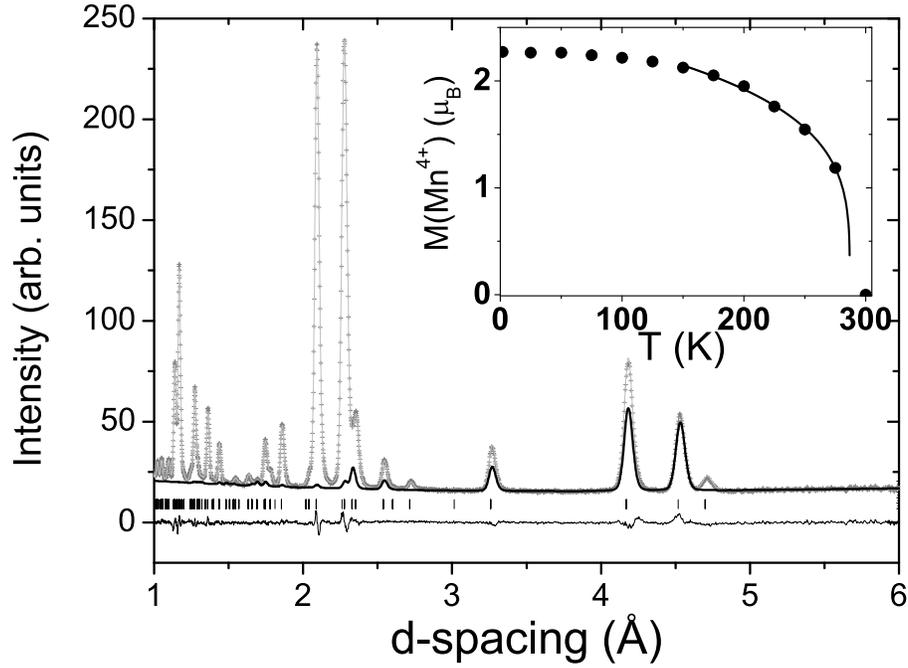}\\
  \caption{Rietveld refinement of \srmno \ at $T=1.5K$ with GEM data collected.
  Data from a detector bank situated at 35$^{\circ}$
  2$\theta$ are shown. The experimental data points are shown as
  grey crosses while the calculated pattern is represented by a grey solid line. Contribution from magnetic
  scattering is plotted separately as a thick black solid line. The lowest continuous line shows the difference curve between data and refinement,
  while tick marks indicate the positions of Bragg reflections.}
  \label{GEMmag}
\end{figure}

\begin{figure}[tbp]
  \includegraphics[width=200pt]{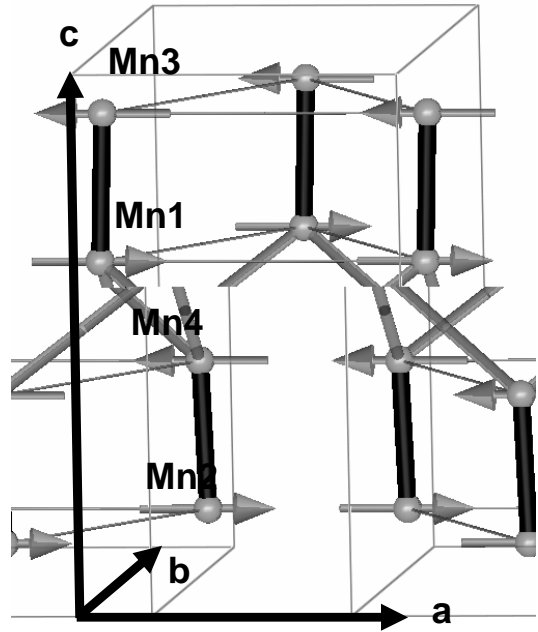}\\
  \caption{Low temperature magnetic structure of \srmno. It is described in the low temperature orthorhombic $C222_1$ cell. The Mn atoms are
  labeled accordingly to the positions of the translated atoms given in Table \ref{magsymm}}
  \label{mag}
\end{figure}

Below room temperature, additional Bragg reflections are observed in
the powder neutron diffraction patterns, which can all be indexed
with a propagation vector \textbf{k}=0 with respect to the
C-centered orthorhombic crystallographic unit cell. These
reflections are clearly magnetic in origin since their intensities
decrease rapidly as a function of the scattering vector Q, following
the magnetic form factor. We note that magnetic scattering and
nuclear scattering attributable to the structural phase transition
give rise to different contributions in the diffraction pattern,
allowing to treat both problems separately. Symmetry analysis of the
magnetic structure has been performed using representation analysis
and is detailed in the appendix \ref{sa}. The magnetic
representation $\Gamma$ decomposed into the direct sum of
irreducible representations:
\begin{equation}
3\Gamma_1+3\Gamma_2+3\Gamma_3+3\Gamma_4
\end{equation}
,where all $\Gamma_i$ (i=1,4) are one-dimensional representations of
the little group G$_k$, which in this case (\textbf{k}=0) is simply
the crystallographic space group. Only two modes, $\psi_7$ or
$\psi_{10}$, belonging to different representations $\Gamma_3$ and
$\Gamma_4$, fit the experimental data. These modes give the same
arrangement of spins with components along the \textit{a}- and
\textit{b}-directions respectively and are in fact equivalent due to
the hexagonal pseudo-symmetry. In fact, these modes belong to a
single irreducible representation when the analysis is performed in
the hexagonal setting. For the same reason, it is impossible to
determine the direction of the spins in the \textit{ab}-plane, and
we have arbitrarily chosen to direct them along the \textit{a}-axis.
The refinement, shown in Fig.\ref{GEMmag}, is of very good quality.
The global user-weigthed $\chi^2=1.30$, and the Bragg R-Factors for
each pattern are $R_{Nuc}=2.44\%$, $R_{Nuc}=2.29\%$,
$R_{Nuc}=2.54\%$, $R_{Mag}=3.09\%$, $R_{Mag}=2.39\%$ and
$R_{Mag}=7.58\%$ for the patterns collected on the $34.96\grad$,
$63.62\grad$, and $91.30\grad$ detectors banks, respectively. The
magnetic structure is reported in Fig.\ref{mag}. It can be described
as a simple antiferromagnetic structure since the magnetic moments
of all first neighbor Mn ions are aligned antiparallel.

The most surprising finding is the value of the manganese magnetic
moment extracted from our Rietveld refinements as a function of
temperature, as shown in Fig.\ref{GEMmag}. For a quenched orbital
contribution, the expected ordered moment for high spin \mnq \ ions
is 3 $\mu_B$ per Mn ion but the observed value, saturating at
2.27(1) at low temperature, is significantly reduced.

\section{Discussion}

\subsection{Structure}

The nature of the structural phase transition and in particular its
connection with the magnetic properties, are not immediately clear.
The transition does not manifest as an abrupt decrease of the
interatomic Mn-Mn distance, so that it cannot be obviously related
to some magnetic exchange-striction, which could strengthen direct
antiferromagnetic interactions over the superexchange interactions.
In the measured temperature range, the Mn-Mn distance decreases
smoothly ($d_{Mn-Mn}$=2.485(2) at 100K and 2.497(3) at 350K)
following essentially the expected thermal expansion. Data collected
on a finer temperature scale would be needed to exclude completely
the presence of exchange striction in this material.

Even though it has not been precisely determined here, the
structural phase transition temperature is certainly above 350K and
it probably coincides with the changes observed in magnetization at
$T_s=$380K. Therefore, it has clearly no connection with the onset
of magnetic order at $T_N=$286K. In fact, preliminary neutron
measurements show that the structural transition temperature
increases when Ca is substituted for Sr in the system, highlighting
the critical role of ionic radius on the A-site. From this, we
conclude that the transition is primarily due to a steric effect.
This probably influences indirectly the magnetic properties as
follows. The distortion induces a small bending of the Mn-O-Mn angle
between some corner sharing octahedra, which could reduce the
electron transfer of the 3d electron between Mn pairs. This can
accordingly enhance correlations associated to the direct exchange
interactions inside each pair below $T_s=380K$, before the magnetic
ordering is established below \tn=286K.

\subsection{Magnetism}

Let's come back first to the exchange topology of \srmno, as
introduced in Fig.\ref{magtopo}. It is clear that for the
superexchange path $J_2$, the nearly 180$\grad$ Mn-O-Mn bond angle
between corner-sharing MnO$_6$ octahedra favors antiferromagnetic
arrangement according to the Goodenough-Kanamori-Anderson (GKA)
rules\cite{goodenough55,Kanamori59}. However, GKA rules predict a
ferromagnetic superexchange interaction for the path $J_1$ between
Mn spins of face-sharing octahedra, since here, the Mn-O-Mn angle is
nearly 90$\grad$. The observed antiferromagnetic arrangement between
these Mn ions indicates that the exchange $J_D$ through direct
overlap of d orbitals, dominates over super-exchange interactions.
The sign of the direct exchange interaction depends on the
interatomic distance between Mn ions, with shorter bonds favoring an
antiferromagnetic arrangement. The effect of interatomic distances
has been systematically studied in Mn intermetallic compounds. In
R(Mn,Fe)$_6$A (R: rare earth, A: Sn or Ge), a crossover from
ferromagnetic to antiferromagnetic interactions was clearly observed
for distances below 2.61 \AA\cite{Marasinghe02}, whereas in various
solid solutions derived from MnSb, that crossover is found for
average Mn-Mn distances below 2.83\AA\cite{Bai84}. A strong
antiferromagnetic direct interaction arising from the small value of
$\sim 2.49\AA$ reported here for the Mn-Mn distance in \srmno\ is
perfectly consistent with the trend seen in intermetallic compounds.

Let us compare now the properties of the present pure, defect free
and stoichiometric \srmno, with that of BaMnO$_3$ and CaMnO$_3$
keeping in mind that most CaMnO$_3$ compounds are oxygen deficient.
BaMnO$_3$\cite{Christensen72} and CaMnO$_3$\cite{Wollan55} have also
simple magnetic structures with AF arrangements of nearest neighbor
Mn. The ordered moment reported for BaMnO$_3$ and that extrapolated
for stoichiometric CaMnO$_3$, fits the expected gS=3$\mu_B$ value
for high spin \mnq \ ions. This is not surprising considering
CaMnO$_3$ as a classical example of Heisenberg magnet with SE
interactions. In the case of BaMnO$_3$, the magnetic ordering shows
a very classical behavior, which probably results from the fact that
the Mn-Mn distance is shorter ($d_{Mn-Mn}=2.40\AA$) than in \srmno.
The enhancement of direct antiferromagnetic exchange interaction in
such case can explain why the moment saturates in BaMnO$_3$ and not
in \srmno, despite the quasi-one dimensional effects expected from
the infinite strings of face sharing octahedra in BaMnO$_3$.

From the presented data and the comparison with other materials, we
therefore suggest that the Heisenberg/SE picture does not fully hold
in \srmno, and that this results from an enhanced competition
between SE and direct exchange interactions between neighboring Mn
atoms in face sharing octahedra. Instead of isolated high spin \mnq
\ ions, the proper magnetic unit should rather be considered as
Mn$_2$O$_9$ pairs. The loss of the \mnq \ ion identity is
reminiscent to the loss of V$^{3+}$ ion identity which has been
discussed in the case of the prototype V$_2$O$_3$
system\cite{Allen76}. Like \srmno, V$_2$O$_3$ shows pairs of face
sharing VO$_6$ octahedra interconnected by the edges in the
corrundum structure. Here also, it has been concluded that the
localized-electron unit in the antiferromagnetic insulating ground
state may be the nearest-neighbor pairs, with the electrons
delocalized within a pair, which has been later thoroughly confirmed
by theoretical investigation of effective spin/orbital
hamiltonians\cite{DiMatteo02}. The reduced moment in \srmno \ at low
temperatures indicates that something similar happens within the
Mn-Mn dimers of \srmno.

\section{Conclusion}

In summary, we have shown that a structural transition does take
place in \srmno, but that this occurs at much higher temperature
than previously reported from Raman scattering\cite{sacchetti05},
and is probably correlated with the transition observed in
magnetization at $\sim 380K$. The structural transformation does not
correspond to a change of polytype, but to a subtle distortion
within the 4L structure lowering the hexagonal $P6_3/mmc$ symmetry
to a $C222_1$ orthorhombic pseudo-symmetry, which preserves the
hexagonal metrics. This structural transition seems to be primarily
due to steric effects, which only indirectly affects magnetic
exchange. NPD reveals that \srmno \  has a reduced moment at low
temperatures. We interpret this might be due to the presence of
delocalized electrons within Mn$_2$O$_9$ pairs, a finding that
deserves confirmation from theoretical studies of the electronic
structure.

\appendix
\section{Symmetry analysis of the structural and magnetic transitions}
\label{sa}

The symmetry properties of the low temperature structure of \srmno \
have been determined by representation analysis. The atomic
positions are written
\begin{equation}
\label{mod}
  \mathbf{r}_{ni}=\mathbf{r}_i^0+\mathbf{u}_{ni}
\end{equation}
for atoms in crystallographic cells indexed by $\mathbf{R}_n$ that
were at average positions $\mathbf{r}_i^0$ in the high temperature
phase, which are now displaced by vectors $\mathbf{u}_{ni}$. These
displacement vectors are decomposed with the fourier sum
\begin{equation}
\label{fourier}
   \mathbf{u}_{ni}=\sum_{\mathbf{k}}\mathbf{u}^i_\mathbf{k}.e^{2i\pi.\mathbf{k}.\mathbf{R}_n}+\mathbf{u}^{i*}_\mathbf{k}.e^{-2i\pi.\mathbf{k}.\mathbf{R}_n}
\end{equation}
where $\mathbf{k}$ is the wave vector characterizing how the
translational symmetry is broken at the transition.

The symmetry relations between the vectors $\mathbf{u}^i_\mathbf{k}$
for atoms of the same crystallographic orbit are obtained for each
Irreducible Representation (Irrep) $\Gamma_\nu$  using the
projection operator:
\begin{equation}
\label{proj}
    \hat{P}^{\nu}=\{\sum_{g\in
    G_\mathbf{k}}D_{\lambda\mu}^{*^\nu}(g)\hat{g}\}
\end{equation}
The sum is over the symmetry elements $g$ of the little group
transforming $\mathbf{k}$ into an equivalent wave vector.
$D_{\lambda\mu}^{*^\nu}(g)$ are elements of the matrix
representation of $g$ for the Irrep $\nu$. This operator projects an
arbitrary set of displacement components on atoms of a
crystallographic orbit of multiplicity $\sigma$, onto symmetrized
basis functions (SBF). We conveniently write these SBF as
$\Psi^\nu_\lambda=\sum^\bigoplus_{i=1..\sigma, \alpha=x,y,z}
a_{i\alpha}.\mathbf{e}_{i\alpha}$, where $\{\mathbf{e}_{i\alpha}\}$
represents the displacement vectors components on an atom $i$ in the
direction $\alpha$. An arbitrary set of linearly independent SBF are
chosen from all the constructed ones. Working out with SBF belonging
to single Irreps is equivalent to assume a distortion compatible
with a second order transition.

In our case, the propagation vector observed experimentally is
\textbf{k}=(0,0,0), a special case where the complex fourier
component $u^i_\mathbf{k}$ identifies with the real displacement
$\mathbf{u}_{ni}$. As always with \textbf{k}=(0 0 0), all the
symmetry elements of $P6_3/mmc$ leave \textbf{k} invariant, hence,
$G_\textbf{k}$ coincides with the space group $P6_3/mmc$. Irreps and
basis functions were obtained using the program BasIreps. In our
case however, the program outputs 8 real one-dimensional
representations, and 4 two-dimensional complex representations. The
SBF being expected to be real, it is desirable to transform the
complex representations into real using an appropriate unitary
similarity transformation matrix $U$, for which
$U^{-1}=(U^*)^T$ before projection. Using \\
\begin{equation}
    U=\frac{1}{2\sqrt{(2-\sqrt{3})^2+1}}\left(
                                          \begin{array}{cc}
                                            2-\sqrt{3} & 1 \\
                                            1 & -2+\sqrt{3} \\
                                          \end{array}
                                        \right)\left(
                                          \begin{array}{cc}
                                            1+i & -1+i \\
                                            1-i & -1-i \\
                                          \end{array}
                                        \right)
\end{equation}
real matrices $d(g)=U^{-1}D(g)U$ for the Irreps
$\Gamma_5$,$\Gamma_6$,$\Gamma_{11}$,$\Gamma_{12}$  have been thus
obtained from the complex matrices $D(g)$, which are output by
BasIreps (see tables II and III).

The projection of the SBF is illustrated for the two-dimensional
representation labeled $\Gamma_5$, which correspond to the Irrep
actually chosen at the phase transition in \srmno. With the special
choice of unitary transform, we could straightforwardly double check
that the symmetry modes associated to the transition from $P6_3/mmc$
to $C222_1$ and given by the program SYMMODES of the Bilbao
crystallographic server\cite{Capillas03} indeed correspond to a
subsets of the SBF calculated for the Irrep $\Gamma_5$ by BasIreps.

Representation analysis is more general, and also applies to other
type of phase transitions where atoms acquire a new scalar,
vectorial, or tensorial property in the low-symmetry phase. Instead
of dealing with atomic displacements (polar vectors) in the case of
structural transitions, we can indeed deal with the appearance of a
magnetic moment to treat the case of magnetic transitions:
\begin{equation}
   \mathbf{\mu}_{ni}=\mathbf{S}^i_\mathbf{k}.e^{2i\pi.\mathbf{k}.R_n}+\mathbf{S}^{i*}_\mathbf{k}.e^{-2i\pi.\mathbf{k}.R_n}
\end{equation}
The only change in the orthogonalization procedure and the
construction of the SBF summarizing this time the symmetry relation
of the fourier components $S^i_\mathbf{k}$, is the action of the
space group operation $g$ in Eq.\ref{proj}, which is different for
axial and polar vectors. We similarly obtained the symmetry analysis
of the possible magnetic structure of \srmno, which is summarized in
table \ref{magsymm}.

\begin{table}[p]
\begin{scriptsize}
\begin{center}
\begin{tabular}[b]{*{7}{c}}
\hline \hline
  Ireps      &         \multicolumn{6}{c}{Symmetry operators}\\
   IT symbol          &  1                     & 3+ 0,0,z              & 2 (0,0,1/2) 0,0,z     & 2 x,x,0               & 3- 0,0,z              & 6- (0,0,1/2) 0,0,z    \\
                      &  $x,y,z          $     & $-x,-y,z+1/2    $     & $-y,x-y,z       $     & $y,-x+y,z+1/2   $     & $-x+y,-x,z      $     & $x-y,x,z+1/2    $     \\
   IT symbol          &  2 0,y,0               & 2 x,-x,1/4            & 6+ (0,0,1/2) 0,0,z    & 2 x,0,0               & 2 2x,x,1/4            & 2 x,2x,1/4            \\
                      &  $y,x,-z         $     & $-y,-x,-z-1/2   $     & $x-y,-y,-z      $     & $-x+y,y,-z-1/2  $     & $-x,-x+y,-z     $     & $x,x-y,-z-1/2   $     \\
   IT symbol          &  -1 0,0,0              & -3+ 0,0,z; 0,0,0      & m x,y,1/4             & m x,-x,z              & -3- 0,0,z; 0,0,0      & -6- 0,0,z; 0,0,1/4    \\
                      &  $-x,-y,-z       $     & $x,y,-z-1/2     $     & $y,-x+y,-z      $     & $-y,x-y,-z-1/2  $     & $x-y,x,-z       $     & $-x+y,-x,-z-1/2 $     \\
   IT symbol          &  m 2x,x,2z             & c x,x,z               & -6+ 0,0,z; 0,0,1/4    & m x,2x,z              & c 0,y,z               & c x,0,z               \\
                      &  $-y,-x,z        $     & $y,x,z+1/2      $     & $-x+y,y,z       $     & $x-y,-y,z+1/2   $     & $x,x-y,z        $     & $-x,-x+y,z+1/2  $     \\
\hline
$\Gamma_1$   &    1      &       1        &       1      &        1    &      1      &       1   \\
             &    1     &        1      &       1       &      1      &       1       &      1   \\
             &    1     &        1    &         1     &        1       &      1      &         1      \\
             &    1     &        1      &       1      &       1      &       1     &        1    \\
$\Gamma_2$   &    1      &       1        &       1      &        1    &      1      &       1   \\
             &    1     &        1      &       1       &      1      &       1       &      1   \\
             &   -1     &       -1    &        -1     &       -1       &     -1      &        -1      \\
             &   -1     &       -1      &      -1      &      -1      &      -1     &       -1    \\
$\Gamma_3$   &    1      &       1        &       1      &        1    &      1      &       1   \\
             &   -1     &       -1      &      -1       &     -1      &      -1       &     -1   \\
             &    1     &        1    &         1     &        1       &      1      &         1      \\
             &   -1     &       -1      &      -1      &      -1      &      -1     &       -1    \\
$\Gamma_4$   &    1      &       1        &       1      &        1    &      1      &       1   \\
             &   -1     &       -1      &      -1       &     -1      &      -1       &     -1   \\
             &   -1     &       -1    &        -1     &       -1       &     -1      &        -1      \\
             &    1     &        1      &       1      &       1      &       1     &        1    \\
$\Gamma_5$   & $ \left( \begin{array}{cc}  1     &  0       \\    0
&  1      \end{array} \right) $  & $ \left( \begin{array}{cc}  1
&  0       \\    0      &  1      \end{array} \right) $  & $ \left(
\begin{array}{cc} -\frac{1}{2} &  \frac{\sqrt{3}}{2}   \\
-\frac{\sqrt{3}}{2}  & -\frac{1}{2}  \end{array} \right) $  & $
\left( \begin{array}{cc} -\frac{1}{2} &  \frac{\sqrt{3}}{2}   \\
-\frac{\sqrt{3}}{2}  & -\frac{1}{2}  \end{array} \right) $  & $
\left( \begin{array}{cc} -\frac{1}{2} & -\frac{\sqrt{3}}{2}   \\
\frac{\sqrt{3}}{2}  & -\frac{1}{2}  \end{array} \right) $  &
$ \left( \begin{array}{cc} -\frac{1}{2} & -\frac{\sqrt{3}}{2}   \\    \frac{\sqrt{3}}{2}  & -\frac{1}{2}  \end{array} \right) $  \\
             &
$ \left( \begin{array}{cc} -\frac{1}{2} & -\frac{\sqrt{3}}{2}   \\
-\frac{\sqrt{3}}{2}  &  \frac{1}{2}  \end{array} \right) $  & $
\left( \begin{array}{cc} -\frac{1}{2} & -\frac{\sqrt{3}}{2}   \\
-\frac{\sqrt{3}}{2}  &  \frac{1}{2}  \end{array} \right) $  & $
\left( \begin{array}{cc}  1     &  0       \\    0      & -1
\end{array} \right) $  & $ \left( \begin{array}{cc}  1     &  0
\\    0      & -1      \end{array} \right) $  & $ \left(
\begin{array}{cc} -\frac{1}{2} &  \frac{\sqrt{3}}{2}   \\
\frac{\sqrt{3}}{2}  &  \frac{1}{2}  \end{array} \right) $  &
$ \left( \begin{array}{cc} -\frac{1}{2} &  \frac{\sqrt{3}}{2}   \\    \frac{\sqrt{3}}{2}  &  \frac{1}{2}  \end{array} \right) $  \\
             &
$ \left( \begin{array}{cc} -1     &  0       \\    0      & -1
\end{array} \right) $  & $ \left( \begin{array}{cc} -1     &  0
\\    0      & -1      \end{array} \right) $  & $ \left(
\begin{array}{cc}  \frac{1}{2} & -\frac{\sqrt{3}}{2}   \\
\frac{\sqrt{3}}{2}  &  \frac{1}{2}  \end{array} \right) $  & $
\left( \begin{array}{cc}  \frac{1}{2} & -\frac{\sqrt{3}}{2}   \\
\frac{\sqrt{3}}{2}  &  \frac{1}{2}  \end{array} \right) $  & $
\left( \begin{array}{cc}  \frac{1}{2} &  \frac{\sqrt{3}}{2}   \\
-\frac{\sqrt{3}}{2}  &  \frac{1}{2}  \end{array} \right) $  &
$ \left( \begin{array}{cc}  \frac{1}{2} &  \frac{\sqrt{3}}{2}   \\   -\frac{\sqrt{3}}{2}  &  \frac{1}{2}  \end{array} \right) $  \\
             &
$ \left( \begin{array}{cc}  \frac{1}{2} &  \frac{\sqrt{3}}{2}   \\
\frac{\sqrt{3}}{2}  & -\frac{1}{2}  \end{array} \right) $  & $
\left( \begin{array}{cc}  \frac{1}{2} &  \frac{\sqrt{3}}{2}   \\
\frac{\sqrt{3}}{2}  & -\frac{1}{2}  \end{array} \right) $  & $
\left( \begin{array}{cc} -1     &  0       \\    0      &  1
\end{array} \right) $  & $ \left( \begin{array}{cc} -1     &  0
\\    0      &  1      \end{array} \right) $  & $ \left(
\begin{array}{cc}  \frac{1}{2} & -\frac{\sqrt{3}}{2}   \\
-\frac{\sqrt{3}}{2}  & -\frac{1}{2}  \end{array} \right) $  &
$ \left( \begin{array}{cc}  \frac{1}{2} & -\frac{\sqrt{3}}{2}   \\   -\frac{\sqrt{3}}{2}  & -\frac{1}{2}  \end{array} \right) $  \\
$\Gamma_6$  & $ \left( \begin{array}{cc}  1     &  0      \\   0
&  1      \end{array}  \right) $   & $ \left( \begin{array}{cc}  1
&  0      \\   0     &  1      \end{array}  \right) $   & $ \left(
\begin{array}{cc} -\frac{1}{2} &  \frac{\sqrt{3}}{2}  \\
-\frac{\sqrt{3}}{2} & -\frac{1}{2}  \end{array}  \right) $   & $
\left( \begin{array}{cc} -\frac{1}{2} &  \frac{\sqrt{3}}{2}  \\
-\frac{\sqrt{3}}{2} & -\frac{1}{2}  \end{array}  \right) $   & $
\left( \begin{array}{cc} -\frac{1}{2} & -\frac{\sqrt{3}}{2}  \\
\frac{\sqrt{3}}{2} & -\frac{1}{2}  \end{array}  \right) $   &
$ \left( \begin{array}{cc} -\frac{1}{2} & -\frac{\sqrt{3}}{2}  \\   \frac{\sqrt{3}}{2} & -\frac{1}{2}  \end{array}  \right) $   \\
     &
$ \left( \begin{array}{cc} -\frac{1}{2} & -\frac{\sqrt{3}}{2}  \\
-\frac{\sqrt{3}}{2} &  \frac{1}{2}  \end{array}  \right) $   & $
\left( \begin{array}{cc} -\frac{1}{2} & -\frac{\sqrt{3}}{2}  \\
-\frac{\sqrt{3}}{2} &  \frac{1}{2}  \end{array}  \right) $   & $
\left( \begin{array}{cc}  1     &  0      \\   0     & -1
\end{array}  \right) $   & $ \left( \begin{array}{cc}  1     &  0
\\   0     & -1      \end{array}  \right) $   & $ \left(
\begin{array}{cc} -\frac{1}{2} &  \frac{\sqrt{3}}{2}  \\
\frac{\sqrt{3}}{2} &  \frac{1}{2}  \end{array}  \right) $   &
$ \left( \begin{array}{cc} -\frac{1}{2} &  \frac{\sqrt{3}}{2}  \\   \frac{\sqrt{3}}{2} &  \frac{1}{2}  \end{array}  \right) $   \\
     &
$ \left( \begin{array}{cc}  1     &  0      \\   0     &  1
\end{array}  \right) $   & $ \left( \begin{array}{cc}  1     &  0
\\   0     &  1      \end{array}  \right) $   & $ \left(
\begin{array}{cc} -\frac{1}{2} &  \frac{\sqrt{3}}{2}  \\
-\frac{\sqrt{3}}{2} & -\frac{1}{2}  \end{array}  \right) $   & $
\left( \begin{array}{cc} -\frac{1}{2} &  \frac{\sqrt{3}}{2}  \\
-\frac{\sqrt{3}}{2} & -\frac{1}{2}  \end{array}  \right) $   & $
\left( \begin{array}{cc} -\frac{1}{2} & -\frac{\sqrt{3}}{2}  \\
\frac{\sqrt{3}}{2} & -\frac{1}{2}  \end{array}  \right) $   &
$ \left( \begin{array}{cc} -\frac{1}{2} & -\frac{\sqrt{3}}{2}  \\   \frac{\sqrt{3}}{2} & -\frac{1}{2}  \end{array}  \right) $   \\
     &
$ \left( \begin{array}{cc} -\frac{1}{2} & -\frac{\sqrt{3}}{2}  \\
-\frac{\sqrt{3}}{2} &  \frac{1}{2}  \end{array}  \right) $   & $
\left( \begin{array}{cc} -\frac{1}{2} & -\frac{\sqrt{3}}{2}  \\
-\frac{\sqrt{3}}{2} &  \frac{1}{2}  \end{array}  \right) $   & $
\left( \begin{array}{cc}  1     &  0      \\   0     & -1
\end{array}  \right) $   & $ \left( \begin{array}{cc}  1     &  0
\\   0     & -1      \end{array}  \right) $   & $ \left(
\begin{array}{cc} -\frac{1}{2} &  \frac{\sqrt{3}}{2}  \\
\frac{\sqrt{3}}{2} &  \frac{1}{2}  \end{array}  \right) $   &
$ \left( \begin{array}{cc} -\frac{1}{2} &  \frac{\sqrt{3}}{2}  \\   \frac{\sqrt{3}}{2} &  \frac{1}{2}  \end{array}  \right) $   \\
\hline \hline
\end{tabular}
\end{center}
\end{scriptsize}
\label{irrep1}
\caption{Representative matrices $d_i(g)$ for each
irreducible representation $\Gamma_i$ of the elements $g$ belonging
to the group of the wave vector $G_\textbf{k}=P6_3/mmc$ for the
space group $G=P6_3/mmc$ and $\textbf{k}=(0 \ 0 \ 0)$}
\end{table}

\begin{table}[p]
\begin{scriptsize}
\begin{center}
\begin{tabular}[b]{*{7}{c}}
\hline \hline
  Ireps      &         \multicolumn{6}{c}{Symmetry operators}\\
   IT symbol          &  1                     & 3+ 0,0,z              & 2 (0,0,1/2) 0,0,z     & 2 x,x,0               & 3- 0,0,z              & 6- (0,0,1/2) 0,0,z    \\
                      &  $x,y,z          $     & $-x,-y,z+1/2    $     & $-y,x-y,z       $     & $y,-x+y,z+1/2   $     & $-x+y,-x,z      $     & $x-y,x,z+1/2    $     \\
   IT symbol          &  2 0,y,0               & 2 x,-x,1/4            & 6+ (0,0,1/2) 0,0,z    & 2 x,0,0               & 2 2x,x,1/4            & 2 x,2x,1/4            \\
                      &  $y,x,-z         $     & $-y,-x,-z-1/2   $     & $x-y,-y,-z      $     & $-x+y,y,-z-1/2  $     & $-x,-x+y,-z     $     & $x,x-y,-z-1/2   $     \\
   IT symbol          &  -1 0,0,0              & -3+ 0,0,z; 0,0,0      & m x,y,1/4             & m x,-x,z              & -3- 0,0,z; 0,0,0      & -6- 0,0,z; 0,0,1/4    \\
                      &  $-x,-y,-z       $     & $x,y,-z-1/2     $     & $y,-x+y,-z      $     & $-y,x-y,-z-1/2  $     & $x-y,x,-z       $     & $-x+y,-x,-z-1/2 $     \\
   IT symbol          &  m 2x,x,2z             & c x,x,z               & -6+ 0,0,z; 0,0,1/4    & m x,2x,z              & c 0,y,z               & c x,0,z               \\
                      &  $-y,-x,z        $     & $y,x,z+1/2      $     & $-x+y,y,z       $     & $x-y,-y,z+1/2   $     & $x,x-y,z        $     & $-x,-x+y,z+1/2  $     \\
\hline
 $\Gamma_7$     &   1        &    -1    &         1      &      -1      &       1      &      -1     \\
                &   1      &      -1      &       1      &      -1    &         1    &        -1       \\
                &   1     &       -1      &       1      &      -1     &        1       &     -1      \\
                &   1     &       -1     &        1     &       -1    &         1     &       -1   \\
 $\Gamma_8$     &   1        &    -1    &         1      &      -1      &       1      &      -1     \\
                &   1      &      -1      &       1      &      -1    &         1    &        -1       \\
                &  -1     &        1      &      -1      &       1     &       -1       &      1      \\
                &  -1     &        1     &       -1     &        1    &        -1     &        1   \\
 $\Gamma_9$     &   1        &    -1    &         1      &      -1      &       1      &      -1     \\
                &  -1      &       1      &      -1      &       1    &        -1    &         1       \\
                &   1     &       -1      &       1      &      -1     &        1       &     -1      \\
                &  -1     &        1     &       -1     &        1    &        -1     &        1   \\
 $\Gamma_{10}$  &   1        &    -1    &         1      &      -1      &       1      &      -1     \\
                &  -1      &       1      &      -1      &       1    &        -1    &         1       \\
                &  -1     &        1      &      -1      &       1     &       -1       &      1      \\
                &   1     &       -1     &        1     &       -1    &         1     &       -1   \\
 $\Gamma_{11}$  &
$ \left( \begin{array}{cc}  1     &  0      \\   0     &  1
\end{array} \right) $   & $ \left( \begin{array}{cc} -1     &  0
\\   0     & -1      \end{array} \right) $   & $ \left(
\begin{array}{cc} -\frac{1}{2} &  \frac{\sqrt{3}}{2}  \\
-\frac{\sqrt{3}}{2} & -\frac{1}{2}  \end{array} \right) $   & $
\left( \begin{array}{cc}  \frac{1}{2} & -\frac{\sqrt{3}}{2}  \\
\frac{\sqrt{3}}{2} &  \frac{1}{2}  \end{array} \right) $   & $
\left( \begin{array}{cc} -\frac{1}{2} & -\frac{\sqrt{3}}{2}  \\
\frac{\sqrt{3}}{2} & -\frac{1}{2}  \end{array} \right) $   &
$ \left( \begin{array}{cc}  \frac{1}{2} &  \frac{\sqrt{3}}{2}  \\  -\frac{\sqrt{3}}{2} &  \frac{1}{2}  \end{array} \right) $   \\
  &   $ \left( \begin{array}{cc} -\frac{1}{2} & -\frac{\sqrt{3}}{2}  \\ -\frac{\sqrt{3}}{2} &  \frac{1}{2}  \end{array} \right) $  &
      $ \left( \begin{array}{cc}  \frac{1}{2} &  \frac{\sqrt{3}}{2}  \\  \frac{\sqrt{3}}{2} & -\frac{1}{2}  \end{array} \right) $  &
      $ \left( \begin{array}{cc}  1     &  0      \\  0     & -1      \end{array} \right) $  &
      $ \left( \begin{array}{cc} -1     &  0      \\  0     &  1      \end{array} \right) $  &
      $ \left( \begin{array}{cc} -\frac{1}{2} &  \frac{\sqrt{3}}{2}  \\  \frac{\sqrt{3}}{2} &  \frac{1}{2}  \end{array} \right) $  &
      $ \left( \begin{array}{cc}  \frac{1}{2} & -\frac{\sqrt{3}}{2}  \\ -\frac{\sqrt{3}}{2} & -\frac{1}{2}  \end{array} \right) $  \\
  &   $ \left( \begin{array}{cc} -1     &  0      \\  0     & -1      \end{array} \right) $  &
      $ \left( \begin{array}{cc}  1     &  0      \\  0     &  1      \end{array} \right) $  &
      $ \left( \begin{array}{cc}  \frac{1}{2} & -\frac{\sqrt{3}}{2}  \\  \frac{\sqrt{3}}{2} &  \frac{1}{2}  \end{array} \right) $  &
      $ \left( \begin{array}{cc} -\frac{1}{2} &  \frac{\sqrt{3}}{2}  \\ -\frac{\sqrt{3}}{2} & -\frac{1}{2}  \end{array} \right) $  &
      $ \left( \begin{array}{cc}  \frac{1}{2} &  \frac{\sqrt{3}}{2}  \\ -\frac{\sqrt{3}}{2} &  \frac{1}{2}  \end{array} \right) $  &
      $ \left( \begin{array}{cc} -\frac{1}{2} & -\frac{\sqrt{3}}{2}  \\  \frac{\sqrt{3}}{2} & -\frac{1}{2}  \end{array} \right) $  \\
  &   $ \left( \begin{array}{cc}  \frac{1}{2} &  \frac{\sqrt{3}}{2}  \\  \frac{\sqrt{3}}{2} & -\frac{1}{2}  \end{array} \right) $  &
      $ \left( \begin{array}{cc} -\frac{1}{2} & -\frac{\sqrt{3}}{2}  \\ -\frac{\sqrt{3}}{2} &  \frac{1}{2}  \end{array} \right) $  &
      $ \left( \begin{array}{cc} -1     &  0      \\  0     &  1      \end{array} \right) $  &
      $ \left( \begin{array}{cc}  1     &  0      \\  0     & -1      \end{array} \right) $  &
      $ \left( \begin{array}{cc}  \frac{1}{2} & -\frac{\sqrt{3}}{2}  \\ -\frac{\sqrt{3}}{2} & -\frac{1}{2}  \end{array} \right) $  &
      $ \left( \begin{array}{cc} -\frac{1}{2} &  \frac{\sqrt{3}}{2}  \\  \frac{\sqrt{3}}{2} &  \frac{1}{2}  \end{array} \right) $  \\
$\Gamma_{12}$   & $ \left( \begin{array}{cc}  1     &  0      \\   0
&  1      \end{array} \right) $   & $ \left( \begin{array}{cc} -1
&  0      \\   0     & -1      \end{array} \right) $   & $ \left(
\begin{array}{cc} -\frac{1}{2} &  \frac{\sqrt{3}}{2}  \\
-\frac{\sqrt{3}}{2} & -\frac{1}{2}  \end{array} \right) $   & $
\left( \begin{array}{cc}  \frac{1}{2} & -\frac{\sqrt{3}}{2}  \\
\frac{\sqrt{3}}{2} &  \frac{1}{2}  \end{array} \right) $   & $
\left( \begin{array}{cc} -\frac{1}{2} & -\frac{\sqrt{3}}{2}  \\
\frac{\sqrt{3}}{2} & -\frac{1}{2}  \end{array} \right) $   &
$ \left( \begin{array}{cc}  \frac{1}{2} &  \frac{\sqrt{3}}{2}  \\  -\frac{\sqrt{3}}{2} &  \frac{1}{2}  \end{array} \right) $   \\
   &  $ \left( \begin{array}{cc} -\frac{1}{2} & -\frac{\sqrt{3}}{2}  \\  -\frac{\sqrt{3}}{2} &  \frac{1}{2}  \end{array} \right) $   &
      $ \left( \begin{array}{cc}  \frac{1}{2} &  \frac{\sqrt{3}}{2}  \\   \frac{\sqrt{3}}{2} & -\frac{1}{2}  \end{array} \right) $   &
      $ \left( \begin{array}{cc}  1     &  0      \\   0     & -1      \end{array} \right) $   &
      $ \left( \begin{array}{cc} -1     &  0      \\   0     &  1      \end{array} \right) $   &
      $ \left( \begin{array}{cc} -\frac{1}{2} &  \frac{\sqrt{3}}{2}  \\   \frac{\sqrt{3}}{2} &  \frac{1}{2}  \end{array} \right) $   &
      $ \left( \begin{array}{cc}  \frac{1}{2} & -\frac{\sqrt{3}}{2}  \\  -\frac{\sqrt{3}}{2} & -\frac{1}{2}  \end{array} \right) $   \\
   &  $ \left( \begin{array}{cc}  1     &  0      \\   0     &  1      \end{array} \right) $   &
      $ \left( \begin{array}{cc} -1     &  0      \\   0     & -1      \end{array} \right) $   &
      $ \left( \begin{array}{cc} -\frac{1}{2} &  \frac{\sqrt{3}}{2}  \\  -\frac{\sqrt{3}}{2} & -\frac{1}{2}  \end{array} \right) $   &
      $ \left( \begin{array}{cc}  \frac{1}{2} & -\frac{\sqrt{3}}{2}  \\   \frac{\sqrt{3}}{2} &  \frac{1}{2}  \end{array} \right) $   &
      $ \left( \begin{array}{cc} -\frac{1}{2} & -\frac{\sqrt{3}}{2}  \\   \frac{\sqrt{3}}{2} & -\frac{1}{2}  \end{array} \right) $   &
      $ \left( \begin{array}{cc}  \frac{1}{2} &  \frac{\sqrt{3}}{2}  \\  -\frac{\sqrt{3}}{2} &  \frac{1}{2}  \end{array} \right) $   \\
   &  $ \left( \begin{array}{cc} -\frac{1}{2} & -\frac{\sqrt{3}}{2}  \\  -\frac{\sqrt{3}}{2} &  \frac{1}{2}  \end{array} \right) $   &
      $ \left( \begin{array}{cc}  \frac{1}{2} &  \frac{\sqrt{3}}{2}  \\   \frac{\sqrt{3}}{2} & -\frac{1}{2}  \end{array} \right) $   &
      $ \left( \begin{array}{cc}  1     &  0      \\   0     & -1      \end{array} \right) $   &
      $ \left( \begin{array}{cc} -1     &  0      \\   0     &  1      \end{array} \right) $   &
      $ \left( \begin{array}{cc} -\frac{1}{2} &  \frac{\sqrt{3}}{2}  \\   \frac{\sqrt{3}}{2} &  \frac{1}{2}  \end{array} \right) $   &
      $ \left( \begin{array}{cc}  \frac{1}{2} & -\frac{\sqrt{3}}{2}  \\  -\frac{\sqrt{3}}{2} & -\frac{1}{2}  \end{array} \right) $   \\
\hline \hline
\end{tabular}
\end{center}
\end{scriptsize}
\label{irrep2} \caption{Continuation of Table II }
\end{table}

\begin{table}[p]
\begin{center}
\begin{tabular}[b]{*{3}{c}}
\hline \hline
\multicolumn{3}{c}{Atom: Sr1$_1$ at position    $(0,0,0)$} \\
\hline
 SYMM     &       x,y,z          &      -x,-y,z+1/2      \\
 Atoms:   &       Sr1$_1$        &         Sr1$_2$       \\
$\Psi_1$  & (     1     0     0) & (    -1     0     0)  \\
$\Psi_2$  & ( -1 -2  0) & (  1  2  0)  \\
\end{tabular}
\\
\begin{tabular}[b]{*{5}{c}}
\hline \hline
\multicolumn{5}{c}{Atom: Mn$_1$  at position    $(\frac{1}{3} \ \frac{2}{3} \ 0.6129)$} \\
\hline
 SYMM     &       x,y,z         &    -x,-y,z+1/2     &        y,x,-z       &    -y,-x,-z+1/2       \\
 Atoms:   &       Mn$_1$        &       Mn$_2$       &        Mn$_3$       &        Mn$_4$         \\
$\Psi_3$  &(     1     0     0) &(    -1     0     0)& (     1     0     0)& (    -1     0     0)  \\
$\Psi_4$  &( -1 -2  0) &(  1  2  0)& ( -1 -2  0)& (  1  2  0)  \\
\end{tabular}
\\
\begin{tabular}[b]{*{7}{c}}
\hline \hline
\multicolumn{7}{c}{Atom: O1$_1$  at position    $(\frac{1}{2} \ 0  \  0)$ } \\
\hline
 SYMM    &         x,y,z        &     -x,-y,z+1/2      &       -y,x-y,z        &   y,-x+y,z+1/2       &    -x+y,-x,z          &      x-y,x,z+1/2      \\
 Atoms:  &        O1$_1$        &        O1$_2$        &         O1$_3$        &        O1$_4$        &         O1$_5$        &         O1$_6$          \\
$\Psi_5$ & (     2     0     0) & (    -2     0     0) &  (     0    -1     0) & (     0     1     0) &  (     1     1     0) &  (    -1    -1     0) \\
$\Psi_6$ & (  0  0  0) & (  0  0  0) &  ( -2 -1  0) & (  2  1  0) &  ( -1  1  0) &  (  1 -1  0) \\
$\Psi_7$ & (  0  0  0) & (  0  0  0) &  (  0  0  1) & (  0  0  1) &  (  0  0 -1) &  (  0  0 -1) \\
$\Psi_8$ & (  0  0  0) & (  0  0  0) &  (  0 -1  0) & (  0  1  0) &  ( -1 -1  0) &  (  1  1  0) \\
$\Psi_9$ & (  0  0  0) & (  0  0  0) &  (  0  1  0) & (  0 -1  0) &  (  1  1  0) &  ( -1 -1  0) \\
$\Psi_{10}$ & (     0     0     2) & (     0     0     2) &  (     0     0    -1) & (     0     0    -1) &  (     0     0    -1) &  (     0     0    -1) \\
\end{tabular}
\\
\begin{tabular}[b]{*{7}{c}}
\hline \hline
\multicolumn{7}{c}{Atom O2$_1$ at position  $( -0.8209 \ 0.8209 \ \frac{3}{4} )$}\\
\hline
 SYMM    &       x,y,z           &     -x,-y,z+1/2      &      -y,x-y,z         &      y,-x+y,z+1/2      &     -x+y,-x,z         &      x-y,x,z+1/2     \\
 Atoms:  &        O2$_1$         &        O2$_2$        &         O2$_3$        &         O2$_4$         &        O2$_5$         &        O2$_6$        \\
$\Psi_{11}$ & (     0     0     1)  & (     0     0     1) &  (     0     0     0) &  (     0     0     0)  & (     0     0    -1)  & (     0     0    -1) \\
$\Psi_{12}$ & (  0  0  1)  & (  0  0  1) &  (  0  0 -2) &  (  0  0 -2)  & (  0  0  1)  & (  0  0  1) \\
\hline \hline
\end{tabular}
\end{center}
\label{bsf} \caption{Basis functions $\psi_i$ of representation
$\Gamma_5$, for each crystallographic site of \srmno.}
\end{table}

\begin{table}[p]
\begin{center}
\begin{tabular}[b]{*{6}{c}}
\hline \hline
\multicolumn{6}{c}{Irreps}          \\
\hline
\multicolumn{2}{c}{IT symbol}       & 1                  &  2 (0,0,1/2) 0,0,z &  2 0,y,1/4         &  2 x,0,0            \\
 \multicolumn{2}{c}{$G_\mathbf{k}$} & $x,y,z$           & $-x,-y,z+1/2$       &  $-x,y,-z+1/2$     &  $x,-y,-z$          \\
\hline
\multicolumn{2}{c}{$\Gamma_1$}   &  1                &        1            &        1           &       1             \\
\multicolumn{2}{c}{$\Gamma_2$}   &  1                &        1            &       -1           &      -1             \\
\multicolumn{2}{c}{$\Gamma_3$}   &  1                &       -1            &        1           &      -1             \\
\multicolumn{2}{c}{$\Gamma_4$}   &  1                &       -1            &       -1           &       1             \\
\hline \hline
\multicolumn{6}{c}{SBF}          \\
\hline
\multicolumn{2}{c}{Atoms}        &      Mn$_1$       &       Mn$_2$        &        Mn$_3$      &        Mn$_4$       \\
\multicolumn{2}{c}{Position}     & $x,y,z$           & $-x,-y,z+1/2$       &  $-x,y,-z+1/2$     &  $x,-y,-z$          \\
\hline
           & $\psi_1$            & $(1  \  0  \  0)$ & $(-1  \  0  \  0 )$ & $(-1  \  0  \  0)$ & $( 1 \   0  \  0)$  \\
$\Gamma_1$ & $\psi_2$            & $(0  \  1  \  0)$ & $( 0  \ -1  \  0 )$ & $( 0  \  1  \  0)$ & $( 0 \  -1  \  0)$  \\
           & $\psi_3$            & $(0  \  0  \  1)$ & $( 0  \  0  \  1 )$ & $( 0  \  0  \ -1)$ & $( 0 \   0  \ -1)$  \\
\hline
           & $\psi_3$            & $(1  \  0  \  0)$ & $(-1  \  0  \  0 )$ & $( 1  \  0  \  0)$ & $(-1 \   0  \  0)$  \\
$\Gamma_2$ & $\psi_4$            & $(0  \  1  \  0)$ & $( 0  \ -1  \  0 )$ & $( 0  \ -1  \  0)$ & $( 0 \   1  \  0)$  \\
           & $\psi_5$            & $(0  \  0  \  1)$ & $( 0  \  0  \  1 )$ & $( 0  \  0  \  1)$ & $( 0 \   0  \  1)$  \\
\hline
           & $\psi_6$            & $(1  \  0  \  0)$ & $( 1  \  0  \  0 )$ & $(-1  \  0  \  0)$ & $(-1 \   0  \  0)$  \\
$\Gamma_3$ & $\psi_7$            & $(0  \  1  \  0)$ & $( 0  \  1  \  0 )$ & $( 0  \  1  \  0)$ & $( 0 \   1  \  0)$  \\
           & $\psi_8$            & $(0  \  0  \  1)$ & $( 0  \  0  \ -1 )$ & $( 0  \  0  \ -1)$ & $( 0 \   0  \  1)$  \\
\hline
           & $\psi_9$            & $(1  \  0  \  0)$ & $( 1  \  0  \  0 )$ & $( 1  \  0  \  0)$ & $( 1 \   0  \  0)$  \\
$\Gamma_4$ & $\psi_10$           & $(0  \  1  \  0)$ & $( 0  \  1  \  0 )$ & $( 0  \ -1  \  0)$ & $( 0 \  -1  \  0)$  \\
           & $\psi_11$           & $(0  \  0  \  1)$ & $( 0  \  0  \ -1 )$ & $( 0  \  0  \  1)$ & $( 0 \   0  \ -1)$  \\
\hline
\end{tabular}
\end{center}
\caption{IRREPS for the little group $G_\mathbf{k}=G$ of G=$C222_1$
and $\mathbf{k}=(0,0,0)$ and SBF for the orbit of the starting atom
Mn$_1$ at position the (-0.0120  0.3333  0.6130).} \label{magsymm}
\end{table}

\bibliographystyle{apsrev}

\begin{thebibliography}{10}
\expandafter\ifx\csname bibnamefont\endcsname\relax
  \def\bibnamefont#1{#1}\fi
\expandafter\ifx\csname bibfnamefont\endcsname\relax
  \def\bibfnamefont#1{#1}\fi
\expandafter\ifx\csname url\endcsname\relax
  \def\url#1{\texttt{#1}}\fi
\expandafter\ifx\csname urlprefix\endcsname\relax\def\urlprefix{URL }\fi
\providecommand{\bibinfo}[2]{#2}
\providecommand{\eprint}[2][]{\url{#2}}

\bibitem{Dagotto01}
\bibinfo{author}{\bibfnamefont{E.}~\bibnamefont{Dagotto}},
  \bibinfo{author}{\bibfnamefont{T.}~\bibnamefont{Hotta}}, \bibnamefont{and}
  \bibinfo{author}{\bibfnamefont{A.}~\bibnamefont{Moreoa}},
  \bibinfo{journal}{Phys. Rep.} \textbf{\bibinfo{volume}{344}},
  \bibinfo{pages}{1} (\bibinfo{year}{2001}).

\bibitem{Orenstein00}
\bibinfo{author}{\bibfnamefont{J.}~\bibnamefont{Orenstein}} \bibnamefont{and}
  \bibinfo{author}{\bibfnamefont{A.}~\bibnamefont{Millis}},
  \bibinfo{journal}{Science} \textbf{\bibinfo{volume}{288}},
  \bibinfo{pages}{468} (\bibinfo{year}{2000}).

\bibitem{Imada98}
\bibinfo{author}{\bibfnamefont{A.}~\bibnamefont{Imada},
  \bibfnamefont{M.~Fujimori}} \bibnamefont{and}
  \bibinfo{author}{\bibfnamefont{T.}~\bibnamefont{Y.}}, \bibinfo{journal}{Rev.
  Mod. Phys.} \textbf{\bibinfo{volume}{70}}, \bibinfo{pages}{1039}
  (\bibinfo{year}{1998}).

\bibitem{goodenough55}
\bibinfo{author}{\bibfnamefont{J.~B.} \bibnamefont{Goodenough}},
  \bibinfo{journal}{Phys. Rev.} \textbf{\bibinfo{volume}{100}},
  \bibinfo{pages}{564} (\bibinfo{year}{1955}).

\bibitem{Anderson59}
\bibinfo{author}{\bibfnamefont{P.}~\bibnamefont{Anderson}},
  \bibinfo{journal}{Phys. Rev.} \textbf{\bibinfo{volume}{115}},
  \bibinfo{pages}{2} (\bibinfo{year}{1959}).

\bibitem{Kanamori59}
\bibinfo{author}{\bibfnamefont{J.}~\bibnamefont{Kanamori}},
  \bibinfo{journal}{J. Phys. Chem. Solids} \textbf{\bibinfo{volume}{10}},
  \bibinfo{pages}{87} (\bibinfo{year}{1959}).

\bibitem{Mason89}
\bibinfo{author}{\bibfnamefont{T.~E.} \bibnamefont{Mason}},
  \bibinfo{author}{\bibfnamefont{B.~D.} \bibnamefont{Gaulin}},
  \bibnamefont{and} \bibinfo{author}{\bibfnamefont{M.~F.}
  \bibnamefont{Collins}}, \bibinfo{journal}{Phys. Rev. B}
  \textbf{\bibinfo{volume}{39}}(\bibinfo{number}{1}), \bibinfo{pages}{586}
  (\bibinfo{year}{1989}).

\bibitem{Harrison91}
\bibinfo{author}{\bibfnamefont{A.}~\bibnamefont{Harrison}},
  \bibinfo{author}{\bibfnamefont{M.~F.} \bibnamefont{Collins}},
  \bibinfo{author}{\bibfnamefont{J.}~\bibnamefont{Abu-Dayyeh}},
  \bibnamefont{and} \bibinfo{author}{\bibfnamefont{C.~V.}
  \bibnamefont{Stager}}, \bibinfo{journal}{Phys. Rev. B}
  \textbf{\bibinfo{volume}{43}}(\bibinfo{number}{1}), \bibinfo{pages}{679}
  (\bibinfo{year}{1991}).

\bibitem{Christensen72}
\bibinfo{author}{\bibfnamefont{A.~N.} \bibnamefont{Christensen}}
  \bibnamefont{and} \bibinfo{author}{\bibfnamefont{G.}~\bibnamefont{Ollivier}},
  \bibinfo{journal}{J. Solid State Chem.} \textbf{\bibinfo{volume}{4}},
  \bibinfo{pages}{131} (\bibinfo{year}{1972}).

\bibitem{Negas70}
\bibinfo{author}{\bibfnamefont{T.}~\bibnamefont{Negas}} \bibnamefont{and}
  \bibinfo{author}{\bibfnamefont{R.~S.} \bibnamefont{Roth}},
  \bibinfo{journal}{J. Solid State Chem.} \textbf{\bibinfo{volume}{1}},
  \bibinfo{pages}{409} (\bibinfo{year}{1970}).

\bibitem{Syono69}
\bibinfo{author}{\bibfnamefont{Y.}~\bibnamefont{Syono}},
  \bibinfo{author}{\bibfnamefont{S.}~\bibnamefont{Akimoto}}, \bibnamefont{and}
  \bibinfo{author}{\bibfnamefont{K.}~\bibnamefont{Kohn}}, \bibinfo{journal}{J.
  Phys. Soc. Jpn.} \textbf{\bibinfo{volume}{26}}, \bibinfo{pages}{993}
  (\bibinfo{year}{1969}).

\bibitem{Chmaissem01}
\bibinfo{author}{\bibfnamefont{O.}~\bibnamefont{Chmaissem}},
  \bibinfo{journal}{Phys. Rev. B} \textbf{\bibinfo{volume}{64}},
  \bibinfo{pages}{134412} (\bibinfo{year}{2001}).

\bibitem{Marasinghe02}
\bibinfo{author}{\bibfnamefont{G.}~\bibnamefont{Marasinghe}},
  \bibinfo{author}{\bibfnamefont{J.}~\bibnamefont{Han}},
  \bibinfo{author}{\bibfnamefont{W.}~\bibnamefont{James}},
  \bibinfo{author}{\bibfnamefont{W.}~\bibnamefont{Yelon}}, \bibnamefont{and}
  \bibinfo{author}{\bibfnamefont{N.}~\bibnamefont{Ali}}, \bibinfo{journal}{J.
  Appl. Phys.} \textbf{\bibinfo{volume}{91}}, \bibinfo{pages}{7863}
  (\bibinfo{year}{2002}).

\bibitem{Bai84}
\bibinfo{author}{\bibfnamefont{V.}~\bibnamefont{Bai}} \bibnamefont{and}
  \bibinfo{author}{\bibfnamefont{T.}~\bibnamefont{Rajasekharan}},
  \bibinfo{journal}{J. of Mag. Mag. Mat.} \textbf{\bibinfo{volume}{42}},
  \bibinfo{pages}{198} (\bibinfo{year}{1984}).

\bibitem{Battle88}
\bibinfo{author}{\bibfnamefont{P.}~\bibnamefont{Battle}},
  \bibinfo{author}{\bibfnamefont{T.}~\bibnamefont{Gibb}}, \bibnamefont{and}
  \bibinfo{author}{\bibfnamefont{C.}~\bibnamefont{Jones}}, \bibinfo{journal}{J.
  Solid State Chem.} \textbf{\bibinfo{volume}{74}}, \bibinfo{pages}{60}
  (\bibinfo{year}{1988}).

\bibitem{sacchetti05}
\bibinfo{author}{\bibfnamefont{A.}~\bibnamefont{Sacchetti}},
  \bibinfo{author}{\bibfnamefont{M.}~\bibnamefont{Baldini}},
  \bibinfo{author}{\bibfnamefont{F.}~\bibnamefont{Crispoldi}},
  \bibinfo{author}{\bibfnamefont{P.}~\bibnamefont{Postorino}},
  \bibinfo{author}{\bibfnamefont{P.}~\bibnamefont{Dore}},
  \bibinfo{author}{\bibfnamefont{A.}~\bibnamefont{Nucara}},
  \bibinfo{author}{\bibfnamefont{C.}~\bibnamefont{Martin}}, \bibnamefont{and}
  \bibinfo{author}{\bibfnamefont{A.}~\bibnamefont{Maignan}},
  \bibinfo{journal}{Phys. Rev. B}
  \textbf{\bibinfo{volume}{72}}(\bibinfo{number}{17}), \bibinfo{pages}{172407}
  (\bibinfo{year}{2005}).

\bibitem{Rietveld69}
\bibinfo{author}{\bibfnamefont{H.}~\bibnamefont{Rietveld}},
  \bibinfo{journal}{J. Appl. Cryst.} \textbf{\bibinfo{volume}{2}},
  \bibinfo{pages}{65} (\bibinfo{year}{1969}).

\bibitem{Rodriguez-Carvajal93}
\bibinfo{author}{\bibfnamefont{J.}~\bibnamefont{Rodr{\'{\i}}guez-Carvajal}},
  \bibinfo{journal}{Physica B} \textbf{\bibinfo{volume}{192}},
  \bibinfo{pages}{55} (\bibinfo{year}{1993}).

\bibitem{Aroyo06}
\bibinfo{author}{\bibfnamefont{M.~I.} \bibnamefont{Aroyo}},
  \bibinfo{author}{\bibfnamefont{J.~M.} \bibnamefont{Perez-Mato}},
  \bibinfo{author}{\bibfnamefont{C.}~\bibnamefont{Capillas}},
  \bibinfo{author}{\bibfnamefont{E.}~\bibnamefont{Kroumova}},
  \bibinfo{author}{\bibfnamefont{S.}~\bibnamefont{Ivantchev}},
  \bibinfo{author}{\bibfnamefont{G.}~\bibnamefont{Madariaga}},
  \bibinfo{author}{\bibfnamefont{A.}~\bibnamefont{Kirov}}, \bibnamefont{and}
  \bibinfo{author}{\bibfnamefont{H.}~\bibnamefont{Wondratschek}},
  \bibinfo{journal}{Z. Kristallogr.}
  \textbf{\bibinfo{volume}{221}}(\bibinfo{number}{1}), \bibinfo{pages}{15}
  (\bibinfo{year}{2006}).

\bibitem{Chamberland70}
\bibinfo{author}{\bibfnamefont{B.~L.} \bibnamefont{Chamberland}},
  \bibinfo{author}{\bibfnamefont{A.~W.} \bibnamefont{Sleight}},
  \bibnamefont{and} \bibinfo{author}{\bibfnamefont{J.~F.}
  \bibnamefont{Weiher}}, \bibinfo{journal}{J. Solid State Chem.}
  \textbf{\bibinfo{volume}{1}}, \bibinfo{pages}{506} (\bibinfo{year}{1970}).

\bibitem{IT83}
\bibinfo{editor}{\bibfnamefont{T.}~\bibnamefont{Hahn}}, ed.,
  \emph{\bibinfo{title}{International Tables for Crystallography, Volume A:
  Space Group Symmetry}} (\bibinfo{publisher}{IUCr, D. Reidel Publishing
  Company}, \bibinfo{year}{1983}).

\bibitem{Stokes84}
\bibinfo{author}{\bibfnamefont{H.~T.} \bibnamefont{Stokes}} \bibnamefont{and}
  \bibinfo{author}{\bibfnamefont{D.~M.} \bibnamefont{Hatch}},
  \bibinfo{journal}{Phys. Rev. B}
  \textbf{\bibinfo{volume}{30}}(\bibinfo{number}{9}), \bibinfo{pages}{4962}
  (\bibinfo{year}{1984}).

\bibitem{Stokes02}
\bibinfo{author}{\bibfnamefont{H.~T.} \bibnamefont{Stokes}} \bibnamefont{and}
  \bibinfo{author}{\bibfnamefont{D.~M.} \bibnamefont{Hatch}},
  \emph{\bibinfo{title}{{ISOTROPY}}} (\bibinfo{year}{2002}),
  \bibinfo{note}{http://stokes.byu.edu/isotropy.html}.

\bibitem{Wollan55}
\bibinfo{author}{\bibfnamefont{E.~O.} \bibnamefont{Wollan}} \emph{et~al.},
  \bibinfo{journal}{Phys. Rev.} \textbf{\bibinfo{volume}{100}},
  \bibinfo{pages}{545} (\bibinfo{year}{1955}).

\bibitem{Allen76}
\bibinfo{author}{\bibfnamefont{J.~W.} \bibnamefont{Allen}},
  \bibinfo{journal}{Phys. Rev. Lett.}
  \textbf{\bibinfo{volume}{36}}(\bibinfo{number}{21}), \bibinfo{pages}{1249}
  (\bibinfo{year}{1976}).

\bibitem{DiMatteo02}
\bibinfo{author}{\bibfnamefont{S.}~\bibnamefont{Di~Matteo}},
  \bibinfo{author}{\bibfnamefont{N.~B.} \bibnamefont{Perkins}},
  \bibnamefont{and} \bibinfo{author}{\bibfnamefont{C.~R.}
  \bibnamefont{Natoli}}, \bibinfo{journal}{Phys. Rev. B}
  \textbf{\bibinfo{volume}{65}}(\bibinfo{number}{5}), \bibinfo{pages}{054413}
  (\bibinfo{year}{2002}).

\bibitem{Capillas03}
\bibinfo{author}{\bibfnamefont{C.}~\bibnamefont{Capillas}},
  \bibinfo{author}{\bibfnamefont{E.}~\bibnamefont{Kroumova}},
  \bibinfo{author}{\bibfnamefont{M.~I.} \bibnamefont{Aroyo}},
  \bibinfo{author}{\bibfnamefont{J.~M.} \bibnamefont{Perez-Mato}},
  \bibinfo{author}{\bibfnamefont{H.~T.} \bibnamefont{Stokes}},
  \bibnamefont{and} \bibinfo{author}{\bibfnamefont{D.~M.} \bibnamefont{Hatch}},
  \bibinfo{journal}{J. Appl. Crystallogr.}
  \textbf{\bibinfo{volume}{36}}(\bibinfo{number}{3 Part 2}),
  \bibinfo{pages}{953} (\bibinfo{year}{2003}).

\end{thebibliography}

\end{document}